# The *Cathaya argyrophylla* Genome Reveals the Evolutionary Trade-offs of a Living Fossil

Yun Wang[1†*], Peng Xie[1†], Shaogang Fan[1†], Zhibo Zhou[1†], Wenyan Zhao[1], Lixuan Xiang[1], Siqin Zhang[1], Lei Sun[4], Ping Mo[1], Xiaolong Jiang[3], Binbin Long[1], Senwei Sun[1], Aihua Deng[1], Haoliang Hu[1,2*], Kerui Huang[1*]

[1]Key Laboratory of Agricultural Products Processing and Food Safety in Hunan Higher Education, Science and Technology Innovation Team for Efficient Agricultural Production and Deep Processing at General University in Hunan Province, Hunan University of Arts and Science, Changde 415000, China.

[2]Institute of Pharmacy and Pharmacology, Hunan Province Cooperative Innovation Center for Molecular Target New Drug Study, Hengyang Medical College, University of South China, Hengyang 421200, China.

[3]College of Forestry, Central South University of Forestry and Technology, Changsha 410000, China.

[4]Chinese Medicinal Materials Breeding Innovation Center of Yuelushan Laboratory, Changsha 410128, China.

## Correspondence:

**1. Yun Wang**, Key Laboratory of Agricultural Products Processing and Food Safety in Hunan Higher Education, Science and Technology Innovation Team for Efficient Agricultural Production and Deep Processing at General University in Hunan Province, Hunan University of Arts and Science, Changde 415000, China.

Email: wy2015@huas.edu.cn

**2. Kerui Huang**, Key Laboratory of Agricultural Products Processing and Food Safety in Hunan Higher Education, Science and Technology Innovation Team for Efficient Agricultural Production and Deep Processing at General University in Hunan Province, Hunan University of Arts and Science, Changde 415000, China.

Email: huangkerui@huas.edu.cn

**3. Haoliang Hu**, Key Laboratory of Agricultural Products Processing and Food Safety in Hunan Higher Education, Science and Technology Innovation Team for Efficient Agricultural Production and Deep Processing at General University in Hunan Province, Hunan University of Arts and Science, Changde 415000, China.

Email: huhl@huas.edu.cn

† These authors contributed equally to this work.

## Abstract
*Cathaya argyrophylla* is an endangered paleoendemic gymnosperm characterized by restricted ecological adaptability and high pathogen susceptibility. To elucidate its genomic architecture and evolutionary history, a *de novo* chromosome-level genome assembly was constructed using PacBio High-Fidelity long reads and Hi-C scaffolding. The resulting 22.73 Gb assembly resolves into 12 pseudochromosomes, demonstrating genome gigantism driven primarily by a 72.92 percent repeat sequence content and extensive intron expansion. Phylogenomic analysis using single-copy orthologs identifies *C. argyrophylla* as a sister lineage to the *Pinus* clade, with an estimated divergence time of 102.8 million years ago. Analysis of gene family dynamics reveals significant expansions in pathways related to membrane lipid metabolism, transmembrane transport, and translation machinery, indicating specific molecular adaptations for cellular homeostasis in resource-limited environments. Conversely, the genome exhibits massive contractions in endogenous defense networks, including plant-pathogen interactions, brassinosteroid signaling, and DNA repair mechanisms. This distinct genomic reduction correlates directly with the slow growth rate and weak innate immunity observed in the species, while the expanded transmembrane transport networks suggest an obligate physiological reliance on symbiotic microbiomes for survival. Ultimately, this reference genome establishes a critical molecular resource for future conservation and breeding programs.

## Introduction

The preservation of paleoendemic species, ancient lineages that survived severe historical climate shifts by retreating into specific, stable refugia, is a window into the evolutionary history of Earth's biodiversity. A classic representative of such relict lineages is the silver fir (*Cathaya argyrophylla*), an endangered gymnosperm endemic to China. Paleobotanical and palynological evidence shows that the ancestors of *Cathaya* were widely distributed across the Northern Hemisphere, including North America, Europe, and Siberia, during the Cretaceous and Tertiary periods[1-3]. However, subsequent climatic deterioration and Quaternary glaciations led to a massive range collapse, which restricted the surviving populations to isolated, mountainous refugia in subtropical China[4,5]. Today, *C. argyrophylla* faces severe ecological challenges. Extensive field surveys and ecological models show its extremely fragmented distribution, slow growth rate, and poor natural regeneration. It also has a high vulnerability to future climate change, and models show a potential loss of over 80% of its climatically suitable habitat under high-emission scenarios[6-9]. Besides, the species shows extreme susceptibility to soil-borne and foliar fungal pathogens[10,11]. For example, it seems to rely heavily on symbiotic microbiomes, such as ectomycorrhizal fungi and endophytic bacteria, for nutrient acquisition and disease resistance[12-14]. While these macroscopic ecological and pathological phenotypes are well-documented, the underlying molecular mechanisms that influence the restricted adaptability and survival strategies of this species are unknown.

Previous studies in gymnosperm genomics show that conifer genomes are quite giant and complex. This is primarily driven by the slow, continuous accumulation of long terminal repeat (LTR) retrotransposons and massive intron expansion, rather than recent whole-genome duplications[15-18]. Recently, researchers applied third-generation long-read sequencing and chromosome conformation capture (Hi-C) technologies to get chromosome-level assemblies for several key gymnosperms, such as *Pinus tabuliformis*[19] and *Ginkgo biloba*[20]. These genomic milestones have advanced our understanding of seed plant evolution, and they show how lineage-specific gene family dynamics affect ecological adaptation. For example, the massive expansion of genes involved in chemical defense and pathogen interaction helps the widespread ecological dominance of the *Pinus* genus[19]. However, the lack of a high-quality reference genome for *C. argyrophylla* is still a gap in comparative gymnosperm genomics. Previously, efforts to determine its phylogenetic position relied on limited morphological traits or organellar markers (such as chloroplast and mitochondrial DNA). While this confirmed its deep evolutionary isolation and general affinity to the *Pinus* clade, it lacked the resolution provided by whole-genome nuclear data[21,22]. More importantly, without a complete nuclear genome, we cannot know if the current ecological fragility of *C. argyrophylla* is caused by genetic erosion, or how its genomic architecture adapts to sustain its survival in isolated refugia over millions of years.

To bridge this knowledge gap and provide a molecular baseline for its conservation[23], we report the *de novo* chromosome-level assembly of the *C. argyrophylla* genome. We integrated PacBio High-Fidelity (HiFi) long reads and Hi-C technologies, which helped us overcome the big assembly challenges posed by this ultra-large and highly repetitive gymnosperm genome. Our primary objectives are to provide a structural annotation, resolve the evolutionary phylogenetic position of *Cathaya* using strictly filtered single-copy orthologs, and explore the changes in its gene repertoire. We expect to find how macroscopic ecological traits link to microscopic genomic features, and whether intron expansion driven by transposable elements influences its genome gigantism. We also want to see how lineage-specific gene family turnovers reflect an evolutionary adaption strategy. We expect that this genomic resource will help to show the molecular basis of the species' ecological vulnerabilities, and provide genetic targets to guide global *ex situ* conservation and breeding programs for this living fossil.

## Materials and Methods

### Plant Materials, Nucleic Acid Extraction, and Sequencing

Fresh needle leaves of *Cathaya argyrophylla* (Sample ID: C2-8Y) were collected from Langshan Rare Plant Research Institute (Xinning County, Hunan) for nucleic acid extraction. High-quality genomic DNA was isolated using an optimized CTAB method. We checked the integrity and purity of the extracted DNA with 0.75% agarose gel electrophoresis and a NanoDrop One spectrophotometer. The DNA concentration was measured using a Qubit 3.0 fluorometer. To do the genome survey and short-read error correction, paired-end sequencing libraries with an insert size of 150 bp were constructed using the Nextera DNA Flex Library Prep Kit and sequenced on the Illumina NovaSeq 6000 and DNBSEQ-T7 platforms[24].

Because the *C. argyrophylla* genome has a huge size and high repeat content, we used third-generation single-molecule real-time (SMRT) sequencing to overcome the assembly challenges. High-molecular-weight DNA was sheared, and a PCR-free SMRTbell library was prepared after damage repair and sequenced on the PacBio Revio platform. For chromosome-level scaffolding, a Hi-C library was built from fresh cross-linked chromatin in situ. The chromatin was digested with the MboI restriction enzyme, filled in with biotin-labeled nucleotides, and ligated. Then, it was sequenced on the Illumina platform to generate paired-end reads.

Besides, to help predict protein-coding genes based on the full-length transcriptome, high-quality total RNA was extracted from the corresponding tissues. Short-read RNA-seq libraries were prepared and sequenced on the Illumina platform. Also, to capture complete transcript isoforms, full-length cDNA libraries were constructed using the SQK-PCS109 kit and sequenced on the Oxford Nanopore PromethION platform[25], along with a PacBio Iso-Seq library.

### Genome Survey and De Novo Assembly

Before assembly, raw short reads from Illumina platforms were processed by fastp (v0.23.2)[26] to remove adapter sequences, poly-N sequences (>5%), and low-quality bases. A genome survey was then conducted to estimate genomic features. The 23-mer frequency distribution of the clean short reads was calculated using Jellyfish (v2.2.10)[27] and evaluated using GenomeScope (v2.0)[28] to estimate the genome size, heterozygosity, and repeat sequence content. To rule out possible exogenous contamination, a subset of 50,000 random clean reads was mapped against the NCBI NT database (version 202107) using BLASTN.

For the primary contig-level assembly, raw PacBio polymerase reads were processed into circular consensus sequencing (CCS) High-Fidelity (HiFi) reads using SMRTLink (v12.0) with filtering parameters (--min-passes=3 and --min-rq=0.99). The HiFi reads were assembled de novo using hifiasm (v0.19.8-r603)[29], an Overlap-Layout-Consensus (OLC) based assembler that helps to resolve haplotype information. Because the genome is very large and heterozygous, purge_dups (v1.2.5)[30] was then used to find and remove redundant haplotypic sequences. This produced a highly contiguous draft contig assembly. The completeness and consistency of the contig assembly were checked by mapping short clean reads back to the genome using BWA (v0.7.17)[31] and running BUSCO (v5.4.7)[32] against the embryophyta_odb10 lineage dataset.

**Hi-C Assisted Chromosome-Level Scaffolding**

To get the draft assembly to the chromosome level, the raw Hi-C reads were preprocessed using fastp and aligned to the contig-level assembly using HiCUP[33]. Invalid interaction read pairs, such as self-circles, dangling ends, and dumped pairs, were filtered out. The valid mapped paired-end reads were used to cluster the contigs into 12 pseudo-chromosomes using the agglomerative hierarchical clustering algorithm in ALLHiC (v0.9.8)[34] (parameter: -e GATC). This is consistent with the documented karyotype of *C. argyrophylla*.

The orientation and ordering of the clustered contigs within each pseudo-chromosome were determined using 3D-DNA (v201008)[35] (parameter: -q 30) and Juicer (v1.6)[36] (parameters: -s MboI -S early). To correct possible inversion and translocation misassemblies, the chromosomal contact maps were visualized and manually adjusted using Juicebox (v1.11.08)[37]. After manual adjustments, adjacent contigs on the pseudo-chromosomes were linked with a gap of 100 'N's. Finally, a genome-wide Hi-C interaction heatmap was generated using HiCExplorer (v3.7.5)[38] to check the accuracy of the anchoring.

**Annotation of Transposable Elements and Non-Coding RNAs**

To understand the massive genome size driven by intron expansion and transposable elements (TEs) in *C. argyrophylla*, we annotated repetitive elements by combining de novo and homology-based strategies. A de novo repeat library was built using RepeatModeler (v2.0.6)[39]. Intact long terminal repeat (LTR) retrotransposons were identified using LTR_FINDER[40] and LTRharvest (v1.6.5)[41]. The intact LTR candidates were integrated and filtered by LTR_retriever (v3.0.1)[42], and unclassified elements were resolved using TEclass (v2.1.3)[43]. The combined de novo library and the RepBase database were used to soft-mask the interspersed TEs across the genome with RepeatMasker (v4.1.7)[44]. Tandem repeats were predicted using Tandem Repeats Finder (TRF, v4.09)[45] and MISA (v2.1).

For non-coding RNAs (ncRNAs), transfer RNAs (tRNAs) and ribosomal RNAs (rRNAs) were identified using tRNAscan-SE (v2.0.12)[46] and RNAmmer (v1.2)[47], respectively. Other structural ncRNAs, like microRNAs (miRNAs) and small nuclear RNAs (snRNAs), were annotated by querying the Rfam database[48] and using the covariance models in INFERNAL (v1.1.4)[49].

**Protein-Coding Gene Prediction and Functional Annotation**

Protein-coding genes were predicted on the repeat-masked genome based on transcriptomic, homology-based, and ab initio evidence. For transcriptomic evidence, short-read RNA-seq reads were aligned using HISAT2 (v2.2.1)[50] and assembled by StringTie (v2.2.1)[51]. Oxford Nanopore full-length reads were filtered by NanoFilt, identified by pychopper, and aligned using minimap2 (v2.26)[52]. Meanwhile, PacBio Iso-Seq data were processed using IsoSeq3 and pbmm2. Transcripts from all sources were merged using TAMA (v1.0)[53], and open reading frames (ORFs) were predicted by TransDecoder (v5.7.0). For homology-based prediction, protein sequences from closely related gymnosperms (*Picea abies*, *Pinus densiflora*, *Pinus lambertiana*, *Pinus tabuliformis*, and *Pinus taeda*) were mapped to the assembly using miniprot (v0.13)[54]. Ab initio prediction was done using Augustus (v3.5.0)[55], Genscan[56], and GlimmerHMM (v3.0.4)[57]. All evidence tracks were combined to make a consensus, high-confidence gene set using EVidenceModeler (v2.1.0)[58] and MAKER (v3.01.03)[59].

To assign functional annotations to the predicted protein-coding genes, we mapped the translated protein sequences against the NCBI Non-Redundant (NR), UniProt, and KOG databases using DIAMOND (v2.1.8)[60](E-value < 1e-5). Protein domains and functional motifs were annotated using hmmscan (v3.3.2)[61]against the Pfam database and InterProScan (v5.55-88.0)[62]. Metabolic pathways were assigned using KofamKOALA[63]to map against the Kyoto Encyclopedia of Genes and Genomes (KEGG) database, and Gene Ontology (GO) terms were extracted.

**Phylogenomic Analysis and Divergence Time Estimation**

To find the evolutionary and phylogenetic position of *C. argyrophylla*, proteomes from 15 representative gymnosperm species were clustered into orthologous gene families using OrthoFinder (v2.5.5)[64]. Single-copy orthologous genes shared across all selected species were extracted. The protein sequences of these single-copy orthologs were aligned using MUSCLE (v3.8.31)[65], and ambiguously aligned regions were trimmed using trimAl (v1.5.rev0)[66]. The concatenated super-matrix was used to build a maximum-likelihood (ML) phylogenetic tree using RAxML (v8.2.13)[67]under the PROTGAMMAWAG substitution model with 100 bootstrap replicates.

Based on this phylogenetic tree, divergence times were estimated using the mcmctree program in the PAML (v4.10.7) package[68]. The molecular clock was calibrated based on fossil records from the TimeTree database. This set temporal constraints on key divergence nodes: *Fokienia hodginsii* vs. *Pinus massoniana* (222.0–298.0 Ma), *Abies alba* vs. *Pinus tabuliformis* (101.1–194.7 Ma), *Pseudotsuga menziesii* vs. *Pinus taeda* (58.5–157.8 Ma), and *Picea glauca* vs. *Picea engelmannii* (1.2–9.9 Ma).

**Gene Family Dynamics, WGD, and Comparative Genomics**

To understand the biological mechanisms for the ecological vulnerability and symbiotic outsourcing of *C. argyrophylla*, the expansion and contraction of gene families were modeled along the time-calibrated tree using a birth-death model in CAFE 5 (v1.1)[69]. A P-value < 0.05 was used as the threshold for significantly expanded or contracted gene families. GO and KEGG functional enrichment analyses were performed on unique, significantly expanded, and significantly contracted gene families using the clusterProfiler R package (v4.2.2)[70]. These analyses showed the changes in intracellular transport, energy metabolism, and the loss of genes in plant-pathogen interactions and DNA repair networks. Positive selection acting on single-copy orthologs was evaluated using the branch-site model in the codeml module of PAML.

To detect macrosynteny and potential whole-genome duplication (WGD) events, an all-against-all protein sequence alignment was performed using BLASTP. Collinear blocks and syntenic gene pairs between *C. argyrophylla* and related species were found using MCScanX (v0.8)[71]and visualized using JCVI (v1.4.23)[72]. Finally, a genomic Circos plot was made using the circlize R package (v0.4.16)[73]. This plot shows chromosome length, gene density, repeat distribution, GC content, and intragenomic collinear links.

## Results

**Genome Assembly, Scaffolding, and Structural Annotation**

To study the genomic characteristics of the silver fir (*Cathaya argyrophylla*), we first did a genome survey using Illumina short-read sequencing data. Based on 23-mer frequency distribution analysis, the genome size was estimated to be approximately 22.57 Gb. It has a quite high repeat sequence content of 72.92% and a 1.47% heterozygosity rate (Figure 1A). To overcome the assembly challenges of this large genome, we used PacBio High-Fidelity (HiFi) long reads to build a *de novo* contig-level assembly. The initial assembly had 11,044 contigs with a total length of 22.73 Gb, which closely matches the K-mer-based estimation. This assembly showed quite good continuity for a giant gymnosperm genome, and achieved a contig N50 of 174.29 Mb and a contig N90 of 29.23 Mb. The maximum contig length reached 603.33 Mb (Table 1). The cumulative contig length distribution curve visually confirmed the robustness of this assembly (Figure 1B). To elevate this assembly to a chromosome scale, we used high-throughput chromosome conformation capture (Hi-C) technology. We mapped valid Hi-C interaction read pairs. About 21.07 Gb of contig sequences, which accounts for 92.68% of the total assembly length, were clustered, ordered, and anchored onto 12 distinct pseudochromosomes. The unplaced sequences were relatively limited, leaving only 1.66 Gb (Table 1). The Hi-C interaction heatmap confirmed this anchoring. It showed strong intra-chromosomal contact signals mostly concentrated along the diagonal line within each of the 12 pseudochromosomes, and background noise among non-homologous chromosomes was quite limited (Figure 1C). This 12-pseudochromosome architecture aligns with previous cytological karyotype reports for the species[74]. After Hi-C scaffolding, the final scaffold N50 length increased to 1.79 Gb, the scaffold N90 reached 1.37 Gb, and the longest pseudochromosome spanned up to 2.01 Gb (Table 1). We integrated the genomic features of the 12 pseudochromosomes and showed them in a Circos plot (Figure 1D).

**Table 1.** Summary statistics of the *Cathaya argyrophylla* genome assembly at the contig and chromosome levels.

| Indexes | Contigs | Chromosomes | Unplaced | Final assembly |
|---|---|---|---|---|
| Number of sequences | 11,044 | 12 | 10,852 | 10,864 |
| Total length (bp) | 22,730,687,697 | 21,066,702,115 | 1,664,003,782 | 22,730,705,897 |
| N50 (bp) | 174,287,548 | 1,801,843,838 | 716,952 | 1,791,507,616 |
| N90 (bp) | 29,229,633 | 1,561,502,637 | 47,649 | 1,374,544,153 |
| Longest (bp) | 603,331,311 | 2,007,795,163 | 35,869,081 | 2,007,795,163 |

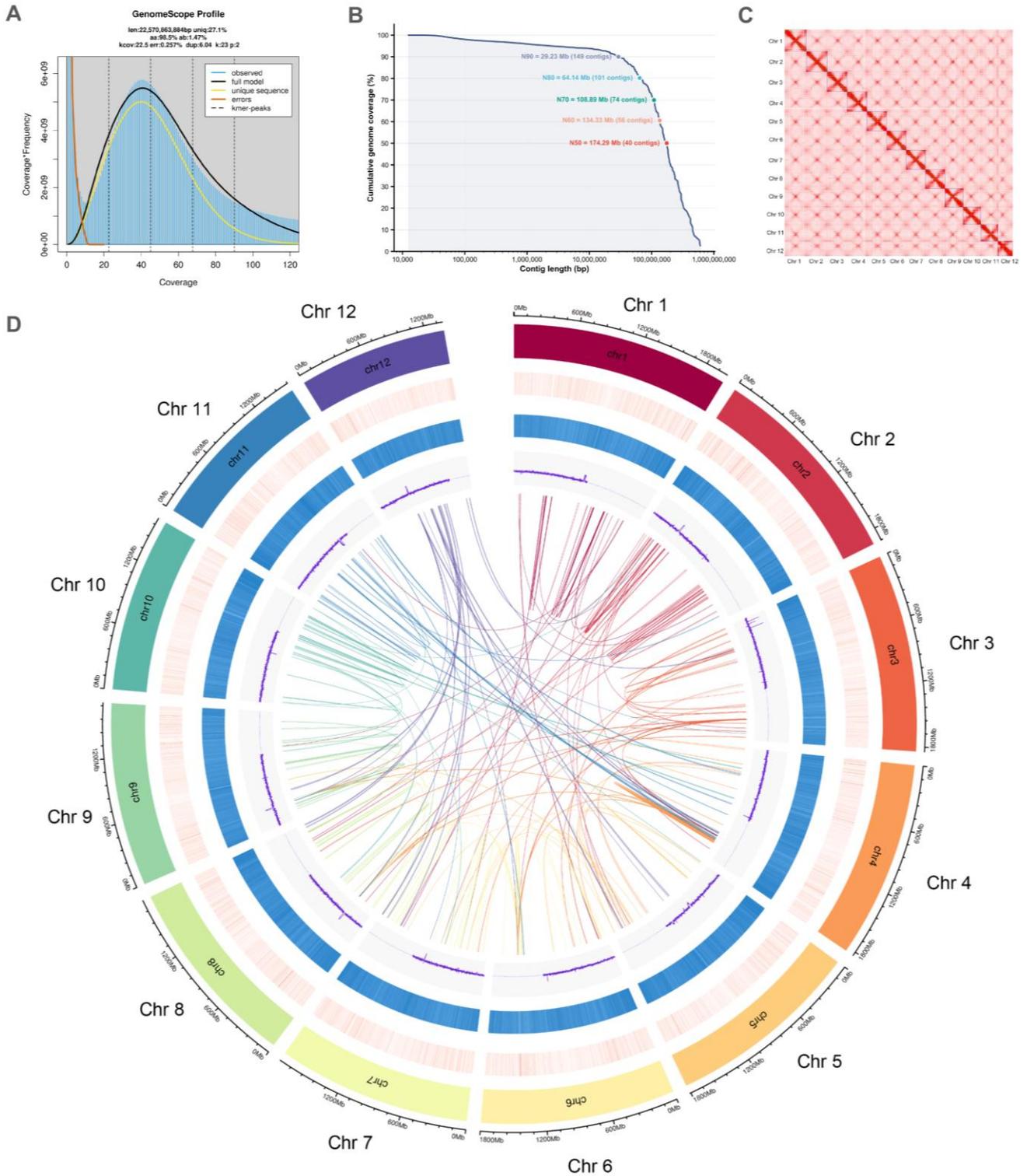

**Figure 1. *De novo* chromosome-level genome assembly and genomic landscape of *Cathaya argyrophylla*.** (A) Genome survey and K-mer analysis. The GenomeScope profile, based on the 23-mer frequency distribution of Illumina short reads, estimates a massive genome size of approximately 22.57 Gb, a high repeat sequence content (72.92%), and a heterozygosity rate of 1.47%. (B) Cumulative contig length distribution curve of the initial *de novo* assembly utilizing PacBio HiFi reads. Highlighted points indicate key continuity metrics ranging from N90 (29.23 Mb) to N50 (174.29 Mb). (C) Genome-wide Hi-C interaction heatmap. The intense intra-chromosomal contact

signals, heavily concentrated along the diagonal line. **(D)** Circos plot visualizing the comprehensive multidimensional genomic features across the 12 pseudochromosomes. The concentric tracks, from the outermost to the innermost, represent: (1) the 12 pseudochromosomes (Chr 1–12) with length scale marks in megabases (Mb); (2) gene density; (3) repeat sequence distribution; (4) GC content; and (5) central colored lines indicating intragenomic collinear links.

After the assembly, we conducted a whole-genome structural annotation. The structural properties of the predicted genes included an average mRNA length of 20,828.49 bp and an average coding sequence (CDS) length of 1,361.71 bp per gene. On average, each gene contained 4.67 exons, and the mean lengths for exons and introns were 384.41 bp and 5,183.78 bp. The total length of the 363,384 identified intronic regions reached approximately 1.88 Gb (Table 2). This massive intron expansion and the 72.92% repeat content provide direct structural evidence for the genome gigantism in this species, which is consistent with the intron expansion driven by transposable elements observed in other giant conifer genomes such as *Pinus tabuliformis*[19] and *Picea abies*[16]. Our annotation pipeline identified 98,972 protein-coding gene models in the *C. argyrophylla* genome (Table 2). This predicted gene count is quite high, and comparative genomic studies on giant gymnosperms show that such numbers often include a large proportion of retroduplications driven by transposable elements, fragmented models, and pseudogenes, as reported in the draft assemblies of *Ginkgo biloba*[20] and *Pinus taeda*[18].

**Table 2.** Statistics of predicted protein-coding genes and structural features in the *Cathaya argyrophylla* genome

| Item | Number |
| --- | --- |
| the total number of gene | 98,972 |
| the average of mRNA_length (bp) | 20,828.49 |
| the average cds_length of per gene (bp) | 1,361.71 |
| the average exon_number of per gene | 4.67 |
| the average of exon_length (bp) | 384.41 |
| the average of intron_length (bp) | 5,183.78 |
| the total number of exon | 462,356 |
| the total number of intron | 363,384 |
| the total intron length (bp) | 1,883,704,030 |

**Comparative Genomics and Phylogenomic Evolution**

To explore structural evolution and chromosomal rearrangements, we performed a macrosynteny analysis across representative lineages using chromosome-level genomes (Table 3). The collinearity blocks between *C. argyrophylla* and three closely related *Pinus* species (*Pinus densiflora*, *P. tabuliformis*, and *P. massoniana*) were quite broad (Figure 2A). This widespread synteny shows that their macro-genomic structures are relatively stable and less prone to large-scale chromosomal rearrangements. This is consistent with the conservative karyotypic evolution widely reported in the Pinaceae family[19,75]. On the other hand, the syntenic comparison with the outgroup species *Fokienia hodginsii* (Cupressaceae) showed narrower collinear blocks and a highly rearranged chromosomal architecture, which reflects the deep evolutionary divergence between these two gymnosperm lineage[76].

**Table 3.** Genomic resources, assembly levels, and data accessions for representative gymnosperm species used in comparative analyses.

| Species | Abbreviation | Assembly Level | Data Source / Accession |
|---|---|---|---|
| *Cathaya argyrophylla* | Carg | Chromosome | Available upon request (This study) |
| *Fokienia hodginsii* | Fhod | Chromosome | https://doi.org/10.6084/m9.figshare.26064412.v1; https://www.ncbi.nlm.nih.gov/bioproject/PRJNA914999 |
| *Abies alba* | Aalb | Scaffold | https://treegenesdb.org/FTP/Genomes/Abal/v1.1/ |
| *Larix kaempferi* | Lkae | Scaffold | https://ngdc.cncb.ac.cn/gwh/Assembly/24470/show |
| *Larix sibirica* | Lsib | Scaffold | https://www.ncbi.nlm.nih.gov/assembly/GCA_004151065.1; https://figshare.com/articles/dataset/Siberian_larch_annotation_repeats_data/19785913/2 |
| *Picea abies* | Pabi | Scaffold | https://treegenesdb.org/FTP/Genomes/Paab/v1.0b; ftp://plantgenie.org/Data/ConGenIE/Picea_abies/v1.0/ |
| *Picea engelmannii* | Peng | Scaffold | https://www.bcgsc.ca/downloads/btl/Spruce/Engelmann_Se404-851/ |
| *Picea glauca* | Pgla | Scaffold | https://www.bcgsc.ca/downloads/btl/Spruce/Pglauca_WS77111/ |
| *Picea mariana* | Pmar | Scaffold | https://doi.org/10.5281/zenodo.7830121 |
| *Picea sitchensis* | Psit | Scaffold | https://www.bcgsc.ca/downloads/btl/Spruce/Psitchensis_Q903/ |
| *Pinus albicaulis* | Palb | Chromosome | https://treegenesdb.org/FTP/temp/P_albicaulis/v1.0/ |
| *Pinus densiflora* | PdeA / PdeB | Chromosome | https://doi.org/10.25452/figshare.plus.25546534 |
| *Pinus lambertiana* | Plam | Scaffold | https://treegenesdb.org/FTP/Genomes/Pila/v1.5/ |
| *Pinus massoniana* | Pmas | Chromosome | http://gigadb.org/dataset/view/id/102688/ |
| *Pinus radiata* | Prad | Scaffold | https://gsajournals.figshare.com/articles/dataset/Supplemental_Material_for_Sturrock_i_et_al_i_2025/29218712?file=55064843 |
| *Pinus tabuliformis* | Ptab | Chromosome | https://www.ncbi.nlm.nih.gov/bioproject/PRJNA784915; https://db.cngb.org/search/project/CNP0001649/; https://figshare.com/articles/dataset/Pinus_tabuliformis_gene_space_annotation/16847146/1 |
| *Pinus taeda* | Ptae | Scaffold | https://treegenesdb.org/FTP/Genomes/Pita/v2.01/ |
| *Pseudotsuga menziesii* | Pmen | Scaffold | https://treegenesdb.org/FTP/Genomes/Psme/v1.0/ |

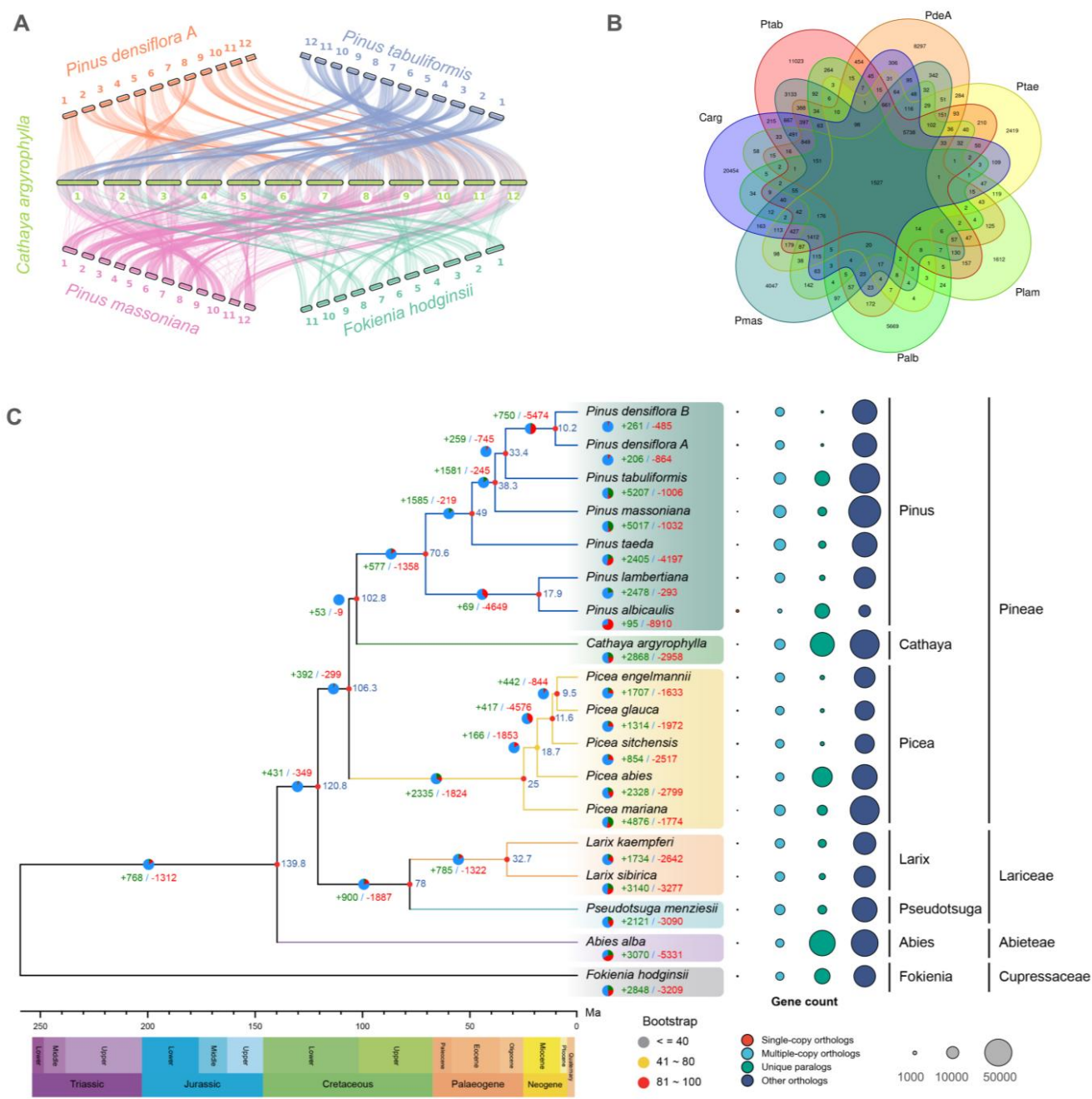

**Figure 2. Comparative genomics, phylogenomic evolution, and gene family dynamics of *Cathaya argyrophylla*.** **(A)** Macrosynteny and collinearity analysis comparing the 12 pseudochromosomes of *C. argyrophylla* (center) with representative chromosome-level gymnosperm genomes. The connecting ribbons denote collinear gene blocks, illustrating broad structural conservation with closely related *Pinus* species (*P. densiflora* A, *P. tabuliformis*, and *P. massoniana*) but a highly rearranged chromosomal architecture relative to the outgroup *Fokienia hodginsii*. **(B)** Venn diagram showing the distribution of shared and unique orthologous gene families among *C. argyrophylla* (Carg) and six representative *Pinus* species (Ptab: *P. tabuliformis*; PdeA: *P. densiflora* A; Ptae: *P. taeda*; Plam: *P. lambertiana*; Palb: *P. albicaulis*; Pmas: *P. massoniana*). **(C)** Time-calibrated maximum-likelihood phylogenomic tree and the distribution of orthologous gene repertoires. The phylogenetic tree (left) was constructed using strictly filtered single-copy orthologous genes. Blue numbers at the nodes indicate the estimated divergence times in millions of years ago (Ma), aligned with the geological time scale at the bottom. The colored circles at the nodes denote gene family dynamics, representing expansion (green), contraction (red), or remaining unchanged (blue). The numbers along the branches provide the exact counts of significantly expanded (green, '+') and contracted (red, '−') gene

families. The bubble plot (right) visualizes the genome-wide statistical breakdown of gene orthology across the analyzed taxa. Gene categories include single-copy orthologs (red), multiple-copy orthologs (cyan), unique paralogs (green), and other orthologs (dark blue), with bubble sizes scaled proportionally to the respective gene counts.

We also looked into the evolutionary changes of the gene repertoire through gene family clustering. A Venn diagram of shared and unique gene families showed that *C. argyrophylla* has a very high number of species-specific gene families, totaling 20,454 (Figure 2B). A statistical breakdown of orthologous gene distribution showed that the *C. argyrophylla* gene repertoire includes 91 single-copy orthologs, 7,097 multiple-copy orthologs, and 54,377 other unclassified orthologs. Besides, it harbors a massive cluster of 37,407 unique paralogous genes, which shows the distinctiveness of its genomic composition compared to other sequenced gymnosperms (Figure 2C). To resolve the evolutionary position of *C. argyrophylla*, we built a maximum-likelihood phylogenomic tree using single-copy orthologous genes. The whole-genome-based phylogenetic topology places *C. argyrophylla* as a sister lineage to the *Pinus* clade (Figure 2C). This whole-genome placement provides evidence that supports previous phylogenetic hypotheses based on limited organellar markers[21,22]. Molecular dating estimations showed that the ancestral branch uniting the *Cathaya* and *Pinus* clades emerged from the *Picea* lineage approximately 106.3 million years ago (Ma), and the divergence between *C. argyrophylla* and the *Pinus* lineage occurred at 102.8 Ma. This timeline based on nuclear data is supported by paleobotanical fossil records from the Cretaceous period[1,3], though it differs from older chloroplast-based estimates of ~144.5 Ma[22]. This reflects the expected cyto-nuclear discordance. Following this split, the *C. argyrophylla* genome underwent gene family turnovers, and 2,868 gene families expanded while 2,958 gene families contracted (Figure 2C).

**Functional Enrichment of Unique and Expanded Gene Families**
To understand the biological role of these lineage-specific genetic changes, we performed Gene Ontology (GO) and Kyoto Encyclopedia of Genes and Genomes (KEGG) enrichment analyses on the unique and expanded gene families. Compared to all other species in the phylogenetic tree, the unique genes of *C. argyrophylla* showed significant functional enrichments in intracellular transport, the endomembrane system, and establishment of localization (Figure 3). For example, there was specific enrichment in the biosynthetic processes of membrane lipid, glycerolipid, and sphingolipid. We also noted a specialization in mitochondrion organization, the mitochondrial inner membrane, and mitochondrial respiratory chain complex I. KEGG analysis also supported this, and it showed enrichments in biotin metabolism, cofactor biosynthesis, fatty acid biosynthesis, unsaturated fatty acid biosynthesis, and oxidative phosphorylation (Figure 4).

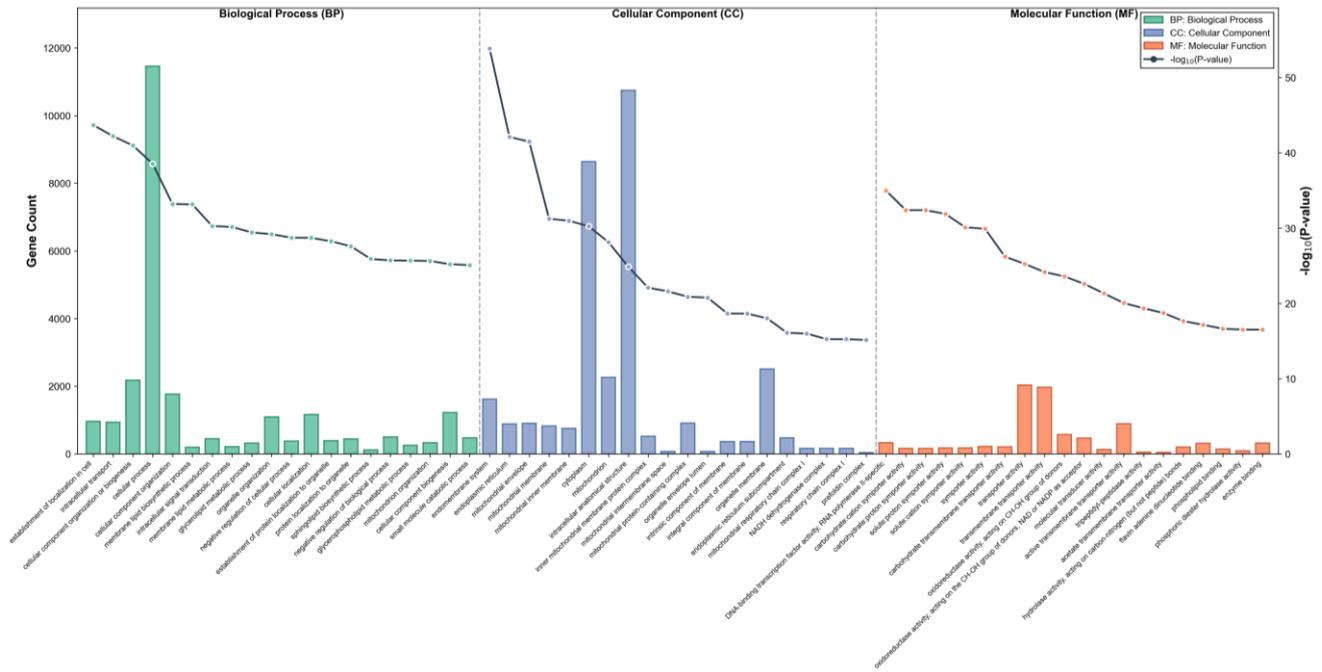

**Figure 3. Gene Ontology (GO) functional enrichment analysis of the unique gene families in the *Cathaya argyrophylla* genome compared across all analyzed lineages.** The composite bar and line chart illustrates the top 20 most significantly enriched GO terms in each of the three primary functional categories: Biological Process (BP, green bars), Cellular Component (CC, blue bars), and Molecular Function (MF, orange bars). The left y-axis and the height of the colored bars represent the absolute number of unique genes (Gene Count) annotated to each specific GO term. The right y-axis and the overlaid line plot with markers indicate the statistical significance of the enrichment for each term, expressed as -$\log_{10}$(P-value).

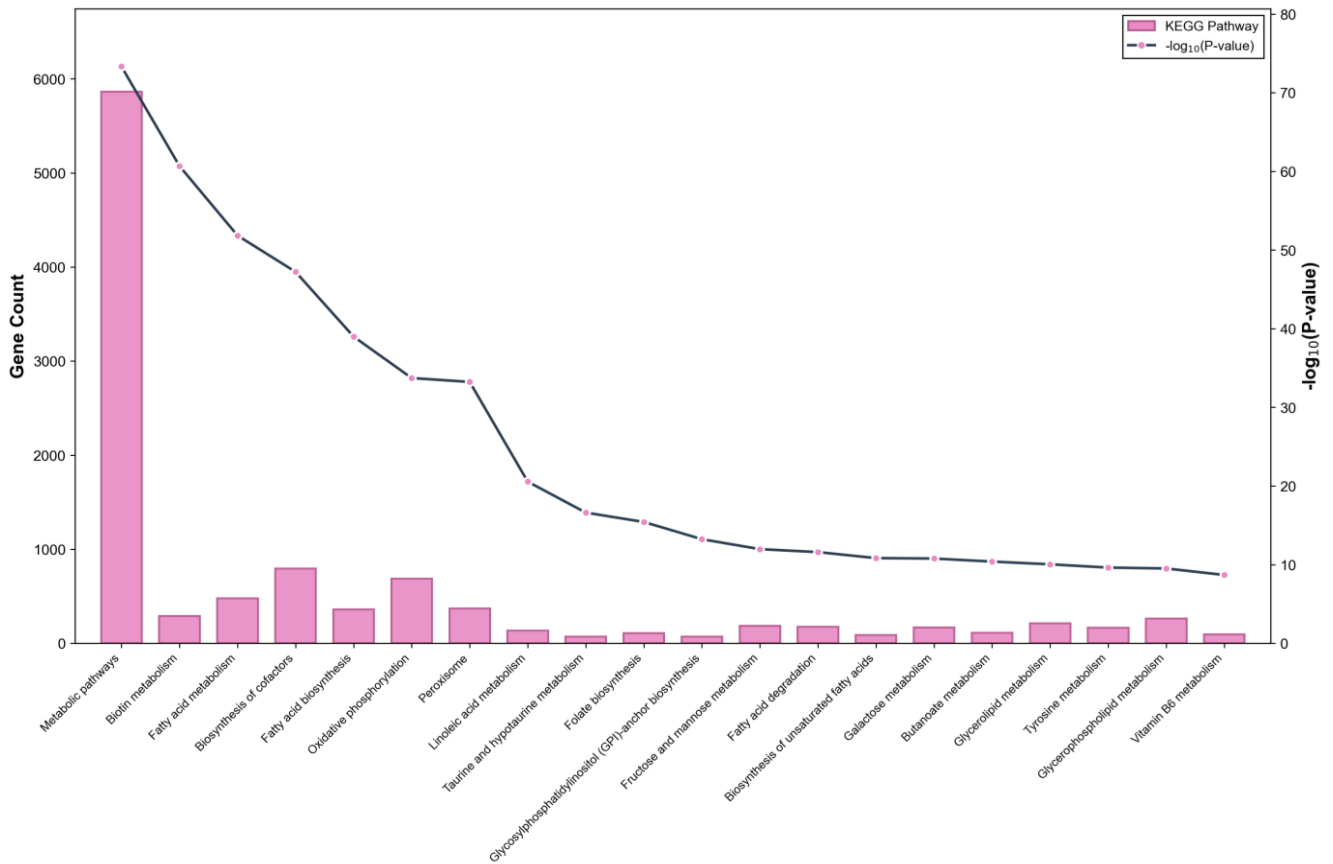

**Figure 4. KEGG pathway enrichment analysis of unique gene families in *Cathaya argyrophylla* compared across all analyzed lineages.** The combined bar and line chart illustrates the top 20 significantly enriched Kyoto Encyclopedia of Genes and Genomes (KEGG) pathways identified from the species-specific unique gene repertoire across all analyzed lineages. The x-axis indicates the names of the enriched KEGG pathways. The pink bars, corresponding to the left y-axis, represent the absolute number of unique genes (Gene Count) assigned to each respective pathway. The solid line with pink markers, corresponding to the right y-axis, denotes the statistical significance of the functional enrichment for each pathway, expressed as -log$_{10}$(P-value). The pathways are arranged from left to right in descending order of their statistical significance, highlighting pronounced metabolic specializations in this species, including biotin metabolism, biosynthesis of cofactors, fatty acid biosynthesis, and oxidative phosphorylation.

When we compared it against its sister genus *Pinus*, the unique genes of *C. argyrophylla* showed enrichment in photosynthesis and light-independent chlorophyll biosynthetic processes, and we found enhanced autophagy and organelle organization (Figure 5). Besides, KEGG enrichments showed porphyrin and chlorophyll metabolism, as well as peroxisome and steroid biosynthesis (Figure 6). Also, the unique gene repertoire had a diversification in transporter activities. The carbohydrate-proton symporters, solute-cation symporters, and ABC transporters played a role here. This expansion of transmembrane transport and lipid metabolism provides a genomic foundation for the species' reliance on ectomycorrhizal symbiotic networks for nutrient acquisition[12,14,77].

**Figure 5. Gene Ontology (GO) functional enrichment analysis of the unique gene families in the *Cathaya argyrophylla* genome relative to all other species in the phylogenetic tree.** The composite bar and line chart illustrates the top 20 most significantly enriched GO terms in each of the three primary functional categories: Biological Process (BP, green bars), Cellular Component (CC, blue bars), and Molecular Function (MF, orange bars). The left y-axis and the height of the colored bars represent the absolute number of unique genes (Gene Count) annotated to each specific GO term. The right y-axis and the overlaid line plot with markers indicate the statistical significance of the enrichment for each term, expressed as -log$_{10}$(P-value).

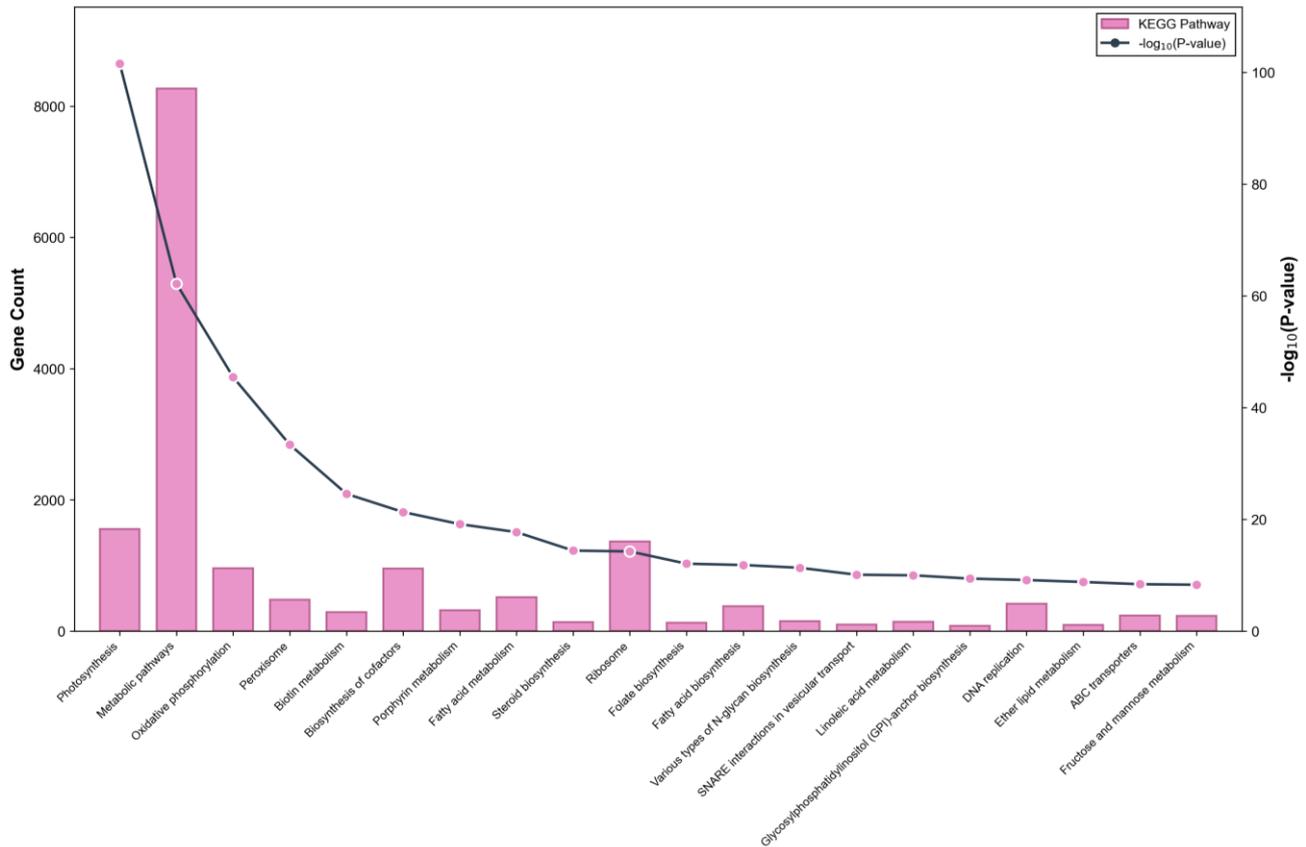

**Figure 6. KEGG pathway enrichment analysis of unique gene families in *Cathaya argyrophylla* relative to all other species in the phylogenetic tree.** The combined bar and line chart illustrates the top 20 significantly enriched Kyoto Encyclopedia of Genes and Genomes (KEGG) pathways identified from the species-specific unique gene repertoire relative to all other species in the phylogenetic tree. The x-axis indicates the names of the enriched KEGG pathways. The pink bars, corresponding to the left y-axis, represent the absolute number of unique genes (Gene Count) assigned to each respective pathway. The solid line with pink markers, corresponding to the right y-axis, denotes the statistical significance of the functional enrichment for each pathway, expressed as -log$_{10}$(P-value). The pathways are arranged from left to right in descending order of their statistical significance, highlighting pronounced metabolic specializations in this species, including biotin metabolism, biosynthesis of cofactors, fatty acid biosynthesis, and oxidative phosphorylation.

Analysis of the expanded gene families showed a functional profile mainly about protein biosynthesis machinery and chloroplast assembly. In the biological process and cellular component categories, the enrichments were related to amide and peptide biosynthetic processes, translation, ribosomes, and structural constituents of ribosomes (Figure 7). Besides, pathways that regulate photosynthesis, the generation of precursor metabolites, plastids, and thylakoid membranes showed massive expansion. KEGG analysis of these expanded families showed a molecular network centered on genetic information processing. It had enrichments in RNA polymerase, spliceosome, aminoacyl-tRNA biosynthesis, and ubiquitin-mediated proteolysis, along with core energy pathways like carbon fixation and the citrate cycle (Figure 8). The expansion of translation machinery and light-independent photosynthetic components suggests a specialized metabolic adaptation. One possibility is that it helps the species to maximize energy capture and cellular homeostasis in the specific, shaded, or resource-limited microhabitats[78,79].

**Figure 7. Gene Ontology (GO) enrichment analysis of expanded gene families in *Cathaya argyrophylla*.** The composite bar and line chart illustrates the top 20 most significantly enriched GO terms in each of the three primary functional categories: Biological Process (BP, green bars), Cellular Component (CC, blue bars), and Molecular Function (MF, orange bars). The left y-axis and the height of the colored bars represent the absolute number of unique genes (Gene Count) annotated to each specific GO term. The right y-axis and the overlaid line plot with markers indicate the statistical significance of the enrichment for each term, expressed as -log₁₀(P-value).

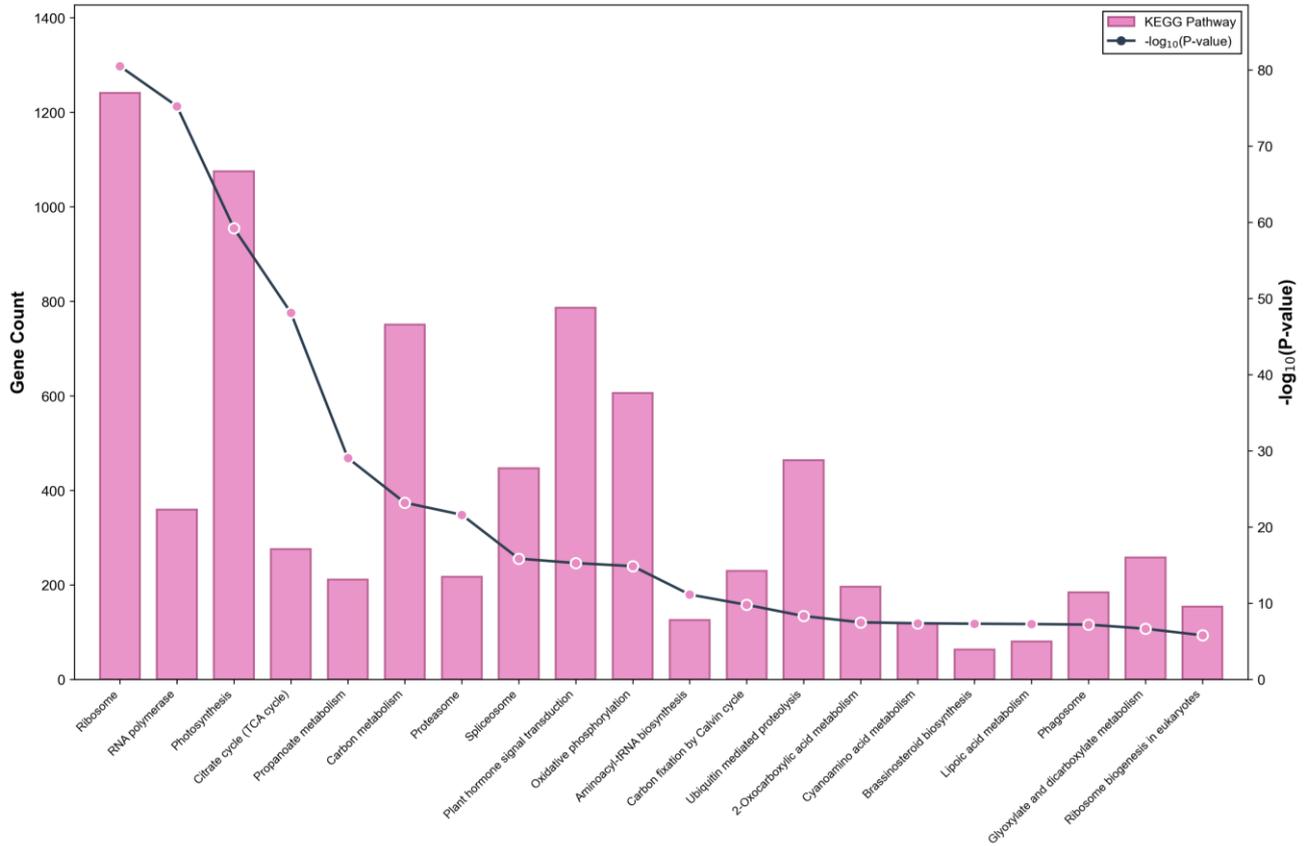

**Figure 8. KEGG pathway enrichment analysis of expanded gene families in *Cathaya argyrophylla*.** The combined bar and line chart illustrates the top 20 significantly enriched Kyoto Encyclopedia of Genes and Genomes (KEGG) pathways identified from the expanded gene families in *Cathaya argyrophylla*. The x-axis indicates the names of the enriched KEGG pathways. The pink bars, corresponding to the left y-axis, represent the absolute number of unique genes (Gene Count) assigned to each respective pathway. The solid line with pink markers, corresponding to the right y-axis, denotes the statistical significance of the functional enrichment for each pathway, expressed as -log$_{10}$(P-value). The pathways are arranged from left to right in descending order of their statistical significance, highlighting pronounced metabolic specializations in this species, including biotin metabolism, biosynthesis of cofactors, fatty acid biosynthesis, and oxidative phosphorylation.

**Functional Enrichment of Contraction genes**

However, the significantly contracted gene families in *C. argyrophylla* showed a "de-redundancy" in environmental signaling and biological defense mechanisms. GO enrichment analysis of the contracted families showed a reduction in phosphorylation processes and the signaling pathway mediated by brassinosteroids. We also found reductions in pH regulation, monovalent inorganic cation homeostasis, and P-type proton/ion transporter activities (Figure 9). Cellular components related to the cell cycle and protein turnover, such as the Cullin-RING and Cul4-RING ubiquitin ligase complexes, and the nuclear replisome, were also contracted. This genomic reduction in growth hormone signaling and DNA replication machinery shows a direct molecular link to the slow growth rate and poor natural regeneration observed in the field[9,80].

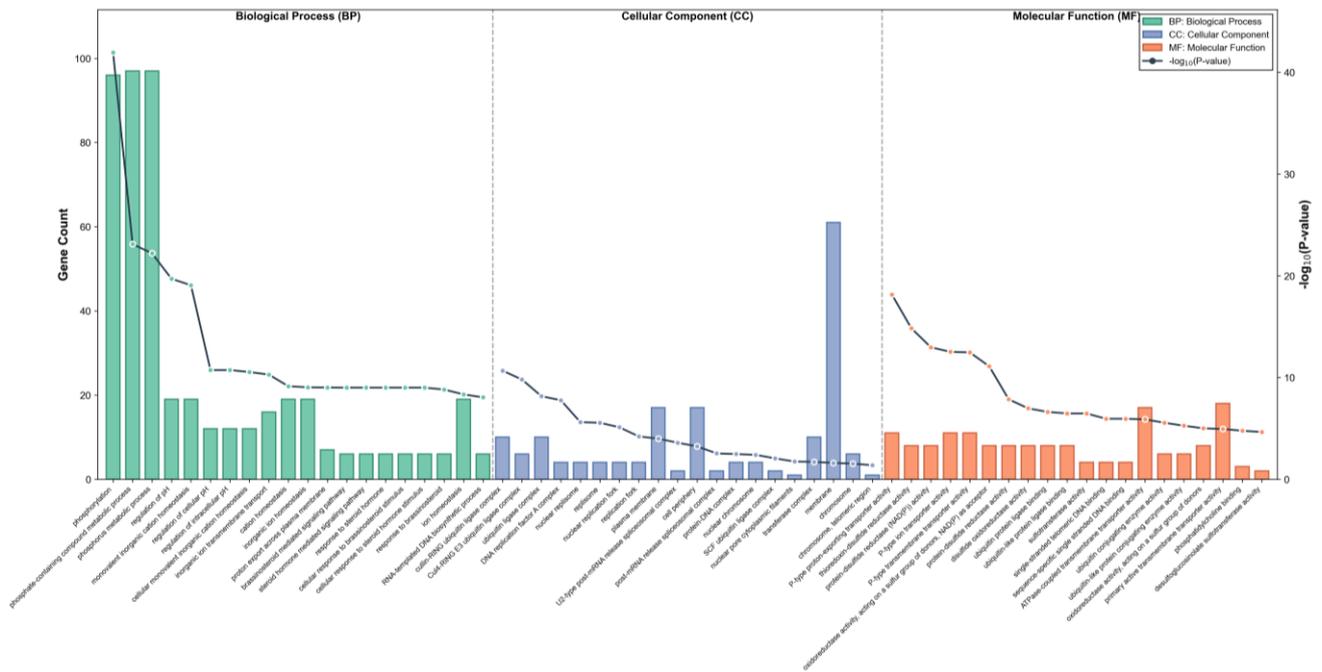

**Figure 9. Gene Ontology (GO) enrichment analysis of contracted gene families in *Cathaya argyrophylla*.** The composite bar and line chart illustrates the top 20 most significantly enriched GO terms in each of the three primary functional categories: Biological Process (BP, green bars), Cellular Component (CC, blue bars), and Molecular Function (MF, orange bars). The left y-axis and the height of the colored bars represent the absolute number of unique genes (Gene Count) annotated to each specific GO term. The right y-axis and the overlaid line plot with markers indicate the statistical significance of the enrichment for each term, expressed as -log$_{10}$(P-value).

Besides, KEGG enrichment analysis of the contracted gene families showed a weakening of the species' biological and chemical defense systems. The pathways for plant-pathogen interaction and the plant MAPK signaling pathway contracted significantly (Figure 10). At the same time, we noted a reduction in the biosynthesis of various plant secondary metabolites. For example, the biosynthesis of flavonoid, terpenoid backbone, and phenylpropanoid decreased. This genomic simplification of immune receptors and chemical defense barriers is quite different from widespread conifers like *Pinus tabuliformis*[19] and *Cycas panzhihuaensis*[17], which have massive expansions in these pathways. The contraction of these defense networks in *C. argyrophylla* correlates with recent reports showing its extreme susceptibility to fungal pathogens. Pathogens like *Fusarium oxysporum* and *Neofusicoccum parvum* affect both natural and cultivated populations and cause rapid mortality[10,11]. Finally, we also found a contraction in DNA repair pathways, including mismatch repair and homologous recombination (Figure 10). The reduction in these repair mechanisms points to genetic vulnerabilities and the risk of accumulating harmful mutations. This is consistent with the species' relict status, historical population bottlenecks, and low population genetic diversity[81,82].

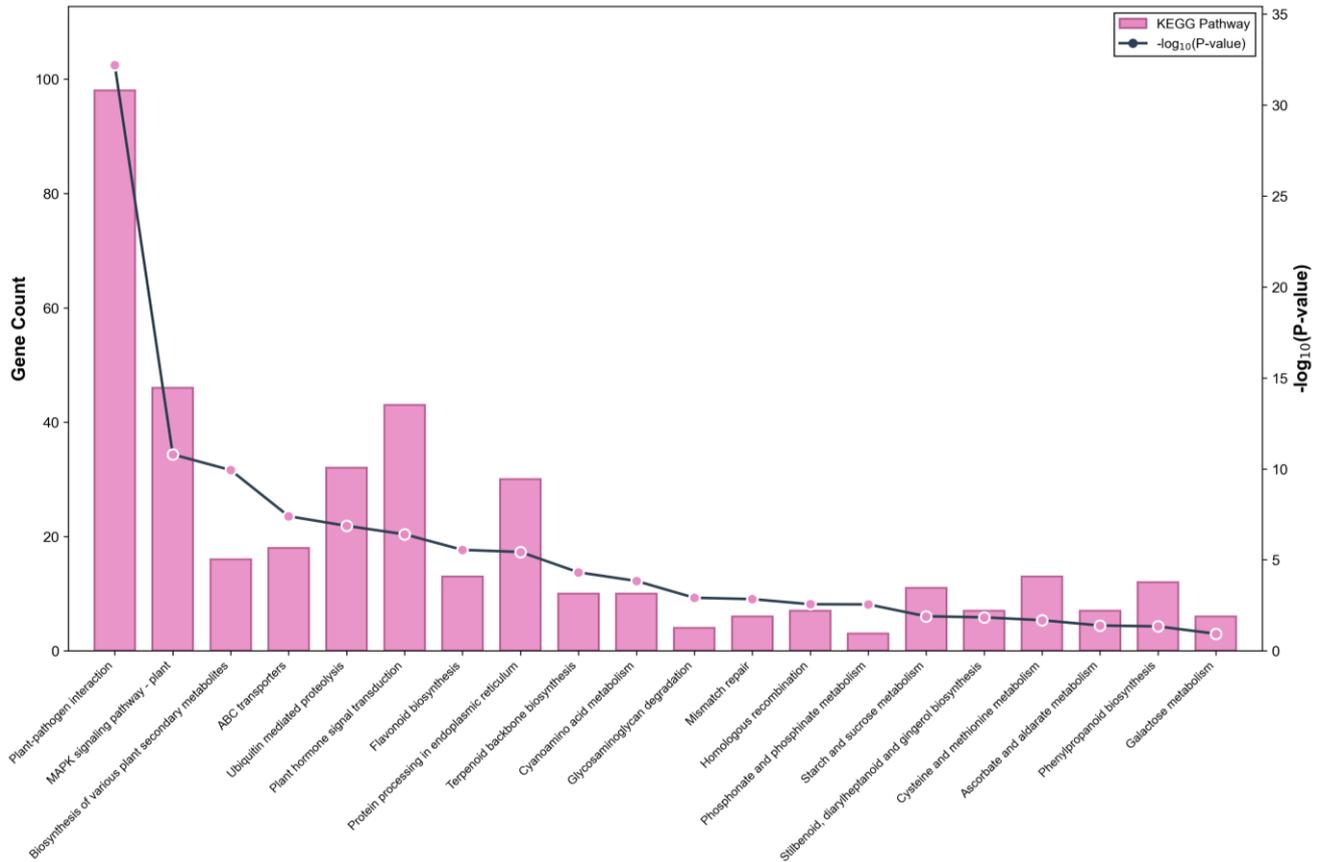

**Figure 10. KEGG pathway enrichment analysis of contracted gene families in *Cathaya argyrophylla*.** The combined bar and line chart illustrates the top 20 significantly enriched Kyoto Encyclopedia of Genes and Genomes (KEGG) pathways identified from the contracted gene families in *Cathaya argyrophylla*. The x-axis indicates the names of the enriched KEGG pathways. The pink bars, corresponding to the left y-axis, represent the absolute number of unique genes (Gene Count) assigned to each respective pathway. The solid line with pink markers, corresponding to the right y-axis, denotes the statistical significance of the functional enrichment for each pathway, expressed as -log$_{10}$(P-value). The pathways are arranged from left to right in descending order of their statistical significance, highlighting pronounced metabolic specializations in this species, including biotin metabolism, biosynthesis of cofactors, fatty acid biosynthesis, and oxidative phosphorylation.

## Discussion

The *de novo* assembly of the *Cathaya argyrophylla* genome provides a molecular baseline for understanding the evolutionary history and ecological fragility of this endangered paleoendemic gymnosperm. We integrated PacBio HiFi long reads and Hi-C scaffolding, overcame the challenges of gymnosperm genome gigantism, and achieved a 22.73 Gb chromosome-level assembly. A remarkably high repeat sequence content (72.92%) and massive intron expansion mainly drive the massive size of the *C. argyrophylla* genome, and the total intronic regions reach 1.88 Gb. This structural architecture aligns with the model of conifer genome evolution, in which the slow and continuous accumulation of long terminal repeat (LTR) retrotransposons within intergenic and intronic regions leads to genome bloat, as seen in *Pinus tabuliformis*[19] and *Picea abies* [16]. Phylogenomically, our whole-genome single-copy ortholog analysis helps to resolve historical taxonomic debates[21]. It places *C. argyrophylla* as a sister lineage to the *Pinus* clade with a divergence time of about 102.8 Ma. This timeline based on nuclear data is supported by Cretaceous fossil pollen records[1-3]. However, it differs from older estimates based on chloroplast DNA (~144.5 Ma)[22]. This shows expected cyto-nuclear discordance. The underlying mechanism may be complex. One possibility is that it comes from differential mutation rates or incomplete lineage sorting during the deep radiation of the Pinaceae family.

Beyond structural architecture, the dynamic turnover of gene families in *C. argyrophylla* provides molecular insights into its current ecological vulnerability. We recently found a massive contraction of gene families related to plant-pathogen interactions, MAPK signaling, brassinosteroid (BR) signaling, and the biosynthesis of secondary metabolites such as flavonoids and terpenoids. This genomic contraction is quite different from widespread conifers like *Pinus tabuliformis*. For example, *P. tabuliformis* shows massive expansions in these chemical defense and immune pathways[19]. The underlying mechanism may be complex. One possibility is that this genomic simplification is a direct consequence of the species' relict history. Having survived Quaternary glaciations by retreating into restricted mountainous refugia ("Plant Museums")[5], *C. argyrophylla* likely experienced relaxed evolutionary pressure to maintain a diverse chemical weapon arsenal. However, this loss of genetic diversity in stress-response networks may explain its modern ecological fragility. The contracted BR signaling and cell cycle machinery may play a role in its slow growth and poor natural regeneration[9,80]. Furthermore, the weakened innate immune system may affect wild and cultivated populations, making them susceptible to soil-borne and foliar fungal pathogens like *Fusarium oxysporum*[10] and *Neofusicoccum parvum*[11]. Besides, we noted a contraction in critical DNA repair mechanisms (mismatch repair and homologous recombination), which suggests a higher risk of accumulating harmful mutations. This is a common feature of ancient lineages with historically small effective population sizes[81-83].

To survive with this contracted endogenous defense system, our data shows that *C. argyrophylla* may have adopted an evolutionary strategy of symbiotic outsourcing, and it relies heavily on the holobiont ecosystem. Ecological and microbiological studies show that *C. argyrophylla* is an obligate host to extensive ectomycorrhizal fungal networks[12,14] and harbors protective root endophytes, such as *Streptomyces cathayae*, which produce secondary metabolites for antibiotic and terpene production[13]. Our genome assembly provides the molecular basis for this symbiosis. The unique and expanded gene families in *C. argyrophylla* are enriched in transmembrane transport, specifically carbohydrate-proton and solute-cation symporters. This suggests an evolutionary trade-off. The plant may have streamlined its own immune and chemical defense networks to tolerate beneficial microbes, and it uses its expanded transporter network to feed carbon to these microbial partners. In exchange, it gets soil nutrient acquisition and biochemical protection against environmental pathogens.

At the same time, the *C. argyrophylla* genome shows metabolic specialization to maximize survival in its specific, resource-limited microhabitats. The massive expansion of gene families related to translation machinery (ribosomes), membrane lipid metabolism (sphingolipids and unsaturated fatty acids), and energy metabolism (oxidative phosphorylation) leads to a highly active basal metabolism. Modulating membrane lipid fluidity is a primary molecular strategy for coping with temperature changes and maintaining cellular homeostasis in harsh mountainous environments[4,84,85]. Also, the unique enrichment of "light-independent chlorophyll biosynthetic processes" and expanded nuclear-encoded photosynthetic genes may be an important adaptation for seedling survival in the deep shade of broad-leaved forests[9]. This nuclear investment in energy metabolism may also be a cyto-nuclear compensatory mechanism. Previous studies show that *C. argyrophylla* has a reduced chloroplast genome lacking *ndh* genes[22,79,86] alongside an 18.99 Mb ultra-large mitochondrial genome[87]. The nuclear genome may have expanded its regulatory repertoire to coordinate with these organellar architectures and maintain efficient energy turnover.

However, there are still some limitations in this study. Relying on a single reference genome means we cannot test population-level structural variations. Future research should use pan-genomic approaches and whole-genome resequencing of individuals from different geographic refugia to quantify mutational load, inbreeding depression, and allele-specific expression. Dual RNA-seq (plant-microbiome transcriptomics) under controlled pathogen stress could also help to test the symbiotic outsourcing of the plant's immune system.

In conclusion, the *C. argyrophylla* genome and its gene family changes affect the survival strategies of this species. The intron expansion driven by TEs, energy and lipid metabolism, and the de-redundancy in its defense and signaling networks play a role in its adaptation. We expect that these genomic findings will provide targets to guide global *ex situ* conservation, habitat modeling, and microbiome-assisted breeding programs for this species[23].

## Acknowledgments
We are grateful to the Langshan Rare Plant Research Institute (Xinning County, Hunan) and the Hunan Botanical Garden for supplying the plant materials that made this work possible. We also appreciate the assistance of their staff during sample collection.